\documentclass[journal,final]{IEEEtran}

\usepackage{cite}      
\usepackage{graphicx}  
\usepackage{psfrag}    
\usepackage{subfigure} 
\usepackage{url}       
\usepackage{stfloats}  
\usepackage{amsmath}   
\usepackage{amssymb}   

\newtheorem{theo}{Theorem}
\newtheorem{corol}{Corollary}

\begin{document}

\title{On the Capacity of the $K$-User Cyclic Gaussian Interference Channel
\thanks{
Manuscript received October 4, 2010; revised May 8, 2012; accepted August
28, 2012. Date of current version August 30, 2012. This work
was supported by the Natural Science and Engineering Research Council (NSERC). The material in this paper was presented
in part at the 2010 IEEE Conference on Information Science and Systems (CISS), and in part at the 2011 IEEE Symposium of Information Theory (ISIT).} \thanks{The authors are with  The Edward S. Rogers Sr. Department of Electrical and Computer Engineering, University of Toronto, Toronto, ON M5S 3G4 Canada (email:
zhoulei@comm.utoronto.ca; weiyu@comm.utoronto.ca). Kindly
address correspondence to Lei Zhou (zhoulei@comm.utoronto.ca).
}\thanks{Copyright (c) 2012 IEEE. Personal use of this material is permitted.  However, permission to use this material for any other purposes must be obtained from the IEEE by sending a request to pubs-permissions@ieee.org.
}
}

\author{ Lei Zhou, {\it Student Member, IEEE} and
	Wei Yu, {\it Senior Member, IEEE}}

\markboth{To appear at IEEE Transactions on Information Theory}
{Zhou and Yu: On the Capacity of the $K$-User Cyclic Gaussian Interference Channel}

\maketitle

\begin{abstract}
This paper studies the capacity region of a $K$-user cyclic Gaussian
interference channel, where the $k$th user interferes with only the
$(k-1)$th user (mod $K$) in the network. Inspired by the work of
Etkin, Tse and Wang, who derived a capacity region outer bound for the
two-user Gaussian interference channel and proved that a simple
Han-Kobayashi power splitting scheme can achieve to within one bit of the
capacity region for all values of channel parameters, this paper shows
that a similar strategy also achieves the capacity region of the
$K$-user cyclic interference channel to within a constant gap in the
weak interference regime. Specifically, for the $K$-user cyclic Gaussian
interference channel, a compact representation of the Han-Kobayashi
achievable rate region using Fourier-Motzkin elimination is first
derived, a capacity region outer bound is then established. It is
shown that the Etkin-Tse-Wang power splitting strategy gives a
constant gap of at most 2 bits in the weak interference regime.  For
the special 3-user case, this gap can be sharpened to $1\frac{1}{2}$
bits by time-sharing of several different strategies.  The
capacity result of the $K$-user cyclic Gaussian interference channel
in the strong interference regime is also given. Further, based on the capacity results, this paper studies the generalized degrees of freedom (GDoF) of the symmetric cyclic interference channel. It is shown that the GDoF of the symmetric capacity is the same as that of the classic two-user interference channel, no matter how many users are in the network.
\end{abstract}


\section{Introduction}
The interference channel models a communication scenario in which several
mutually interfering transmitter-receiver pairs share the same
physical medium. The interference channel is a useful model for many
practical systems such as the wireless network. The capacity region
of the interference channel, however, has not been completely
characterized, even for the two-user Gaussian case.

The largest known achievable rate region for the two-user interference
channel is given by Han and Kobayashi \cite{HK1981} using a coding
scheme involving common-private power splitting. Chong et
al.\ \cite{Chong2006} obtained the same rate region in a simpler form
by applying the Fourier-Motzkin algorithm together with a time-sharing
technique to the Han and Kobayashi's rate region characterization.
The optimality of the Han-Kobayashi region for the two-user Gaussian
interference channel is still an open problem in general, except in the
strong interference regime where transmission with common information
only achieves the capacity region \cite{HK1981, Carleial1978,
Sato}, and in a noisy interference regime where transmission with
private information only achieves the sum capacity \cite{VVV,
Khandani08, Biao}.

In a breakthrough, Etkin, Tse and Wang \cite{Tse2007} showed that the Han-Kobayashi scheme can in fact
achieve to within one bit of the capacity region for the two-user
Gaussian interference channel for all channel parameters. Their key
insight was that the interference-to-noise ratio (INR) of
the private message should be chosen to be as close to $1$ as possible
in the Han-Kobayashi scheme. They also found a new capacity region
outer bound using a genie-aided technique. In the rest of this paper, we refer this particular setting of the private message power as the Etkin-Tse-Wang (ETW) power-splitting strategy.

The Etkin, Tse and Wang's result applies only to the two-user
interference channel. Practical systems often have more
than two transmitter-receiver pairs, yet the generalization of Etkin,
Tse and Wang's work to the interference channels with more than two
users has proved difficult for the following reasons. First, it
appears that the Han-Kobayashi common-private superposition coding is
no longer adequate for the $K$-user interference channel.
Interference alignment types of coding scheme
\cite{Jafar_interference_alignment} \cite{many-to-one} can potentially
enlarge the achievable rate region.  Second, even within the
Han-Kobayashi framework, when more than two receivers are involved,
multiple common messages at each transmitter may be needed, making the
optimization of the resulting rate region difficult.

In the context of $K$-user Gaussian interference channels, sum
capacity results are available in the noisy interference regime
\cite{VVV, Shang_ICC2008}. In particular, Annapureddy et al.\ \cite{VVV} obtained the sum capacity for the symmetric three-user Gaussian
interference channel, the one-to-many, and the many-to-one Gaussian
interference channels under the noisy interference criterion. Similarly, Shang et al.\ \cite{Shang_ICC2008} studied the
fully connected $K$-user Gaussian interference channel and showed that
treating interference as noise at the receiver is sum-capacity
achieving when the transmit power and the cross channel gains are
sufficiently weak
to satisfy a certain criterion. Further, achievability and outer bounds for the three-user interference channel have also been studied in  \cite{3uesr_Daniela} and  \cite{3user_strongverystrong}. Finally, much work has
been carried out on the generalized degree of freedom (GDoF as defined
in \cite{Tse2007}) of the $K$-user interference channel and its
variations \cite{Jafar_interference_alignment, Jafar_gdof_K,
Jafar_gdof_many_to_one, Sum_Sym_K_Nazer}.

\begin{figure} [t]
\centering
\includegraphics[width=3.4in]{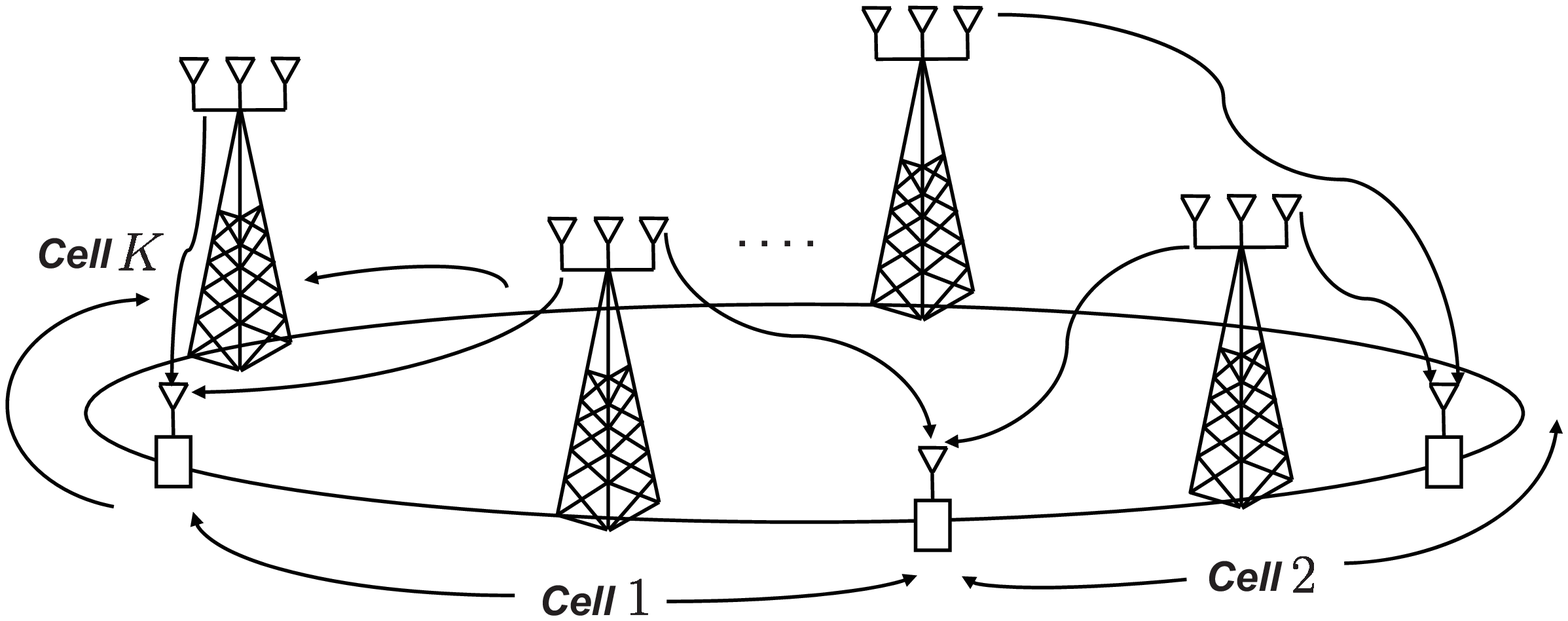}
\caption{The circular array soft-handoff model} \label{modified_wyner_model}
\end{figure}

Instead of treating the general $K$-user interference channel, this
paper focuses on a cyclic Gaussian interference channel, where
the $k$th user interferes with only the $(k-1)$th user. In this case,
each transmitter interferes with only one other receiver, and each
receiver suffers interference from only one other transmitter, thereby
avoiding the difficulties mentioned earlier. For the $K$-user cyclic
interference channel, the Etkin, Tse and Wang's coding strategy remains a natural one. The main objective of this paper is to show that it indeed
achieves to within a constant gap of the capacity region for this cyclic
model in the weak interference regime to be defined later.

The cyclic interference channel model is motivated by the so-called
modified Wyner model, which describes the soft handoff scenario of a
cellular network \cite{Somekh_softhandoff}. The original Wyner model
\cite{Wyner_softhandoff} assumes that all cells are arranged in a
linear array with the base-stations located at the center of each cell,
and where intercell interference comes from only the two adjacent cells.
In the modified Wyner model \cite{Somekh_softhandoff} cells are
arranged in a circular array  as shown in Fig.~\ref{modified_wyner_model}. The  mobile terminals are located along the circular array. If one assumes that the mobile terminals always communicate with the intended base-station to their left (or
right), while only suffering from interference due to the base-station
to their right (or left), one arrives at the $K$-user cyclic Gaussian
interference channel studied in this paper. The modified Wyner model
has been extensively studied in the literature \cite{Somekh_softhandoff, Liang_softhandoff, Sheng_softhandoff}, but often either with interference treated as noise or with the assumption of full base-station cooperation. This paper studies the modified Wyner model without base-station cooperation, in
which case the soft-handoff problem becomes that of a cyclic interference channel.

This paper primarily focuses on the $K$-user cyclic Gaussian
interference channel in the weak interference regime.  The main
contributions of this paper are as follows. This
paper first derives a compact characterization of the Han-Kobayashi
achievable rate region by applying the Fourier-Motzkin elimination
algorithm. A capacity region outer bound is then obtained. It is shown
that with the Etkin, Tse and Wang's coding strategy, one can achieve
to within $1\frac{1}{2}$ bits of the capacity region when $K=3$ (with
time-sharing), and to within two bits of the capacity region in
general in the weak interference regime. Finally, the capacity result
for the strong interference regime is also derived.

A key part of the development involves a Fourier-Motzkin elimination
procedure on the achievable rate region of the $K$-user cyclic
interference channel.
To deal with the large number of inequality constraints, an induction
proof is used.
It is shown that as compared to the two-user case, where the rate
region is defined by constraints on the individual rate $R_i$, the sum
rate $R_1+R_2$, and the sum rate plus an individual rate $2R_i + R_j$
($i \neq j$), the achievable rate region for the $K$-user cyclic
interference channel is defined by an additional set of constraints on
the sum rate of any arbitrary $l$ adjacent users, where $2 \le l < K$.
These four types of rate constraints completely characterize the
Han-Kobayashi region for the $K$-user cyclic interference channel.
They give rise to a total of $K^2+1$ constraints.

For the symmetric
$K$-user cyclic channel where all direct links share the same channel
gain and all cross links share another channel gain, it is shown that the GDoF of the symmetric capacity is not dependent on the number of users in the network. Therefore, adding more users to a $K$-user cyclic interference channel with symmetric channel parameters does not affect the per-user rate. 

\section{Channel Model}

\begin{figure} [t]
\centering
\centering \psfrag{X1}{$X_1$} \psfrag{X2}{$X_2$} \psfrag{X3}{$X_3$} \psfrag{XK}{$X_K$}\psfrag{Y1}{$Y_1$} \psfrag{Y2}{$Y_2$} \psfrag{Y3}{$Y_3$} \psfrag{YK}{$Y_K$} \psfrag{Z1}{$Z_1$} \psfrag{Z2}{$Z_2$} \psfrag{Z3}{$Z_3$} \psfrag{ZK}{$Z_K$} \psfrag{h11}{$h_{11}$} \psfrag{h22}{$h_{22}$} \psfrag{h33}{$h_{33}$} \psfrag{hkk}{$h_{KK}$} \psfrag{h21}{$h_{21}$}  \psfrag{h32}{$h_{32}$}  \psfrag{h1k}{$h_{1K}$}  \psfrag{leftdots}{$\vdots$}  \psfrag{rightdots}{$\vdots$}
\includegraphics[width=2.5in]{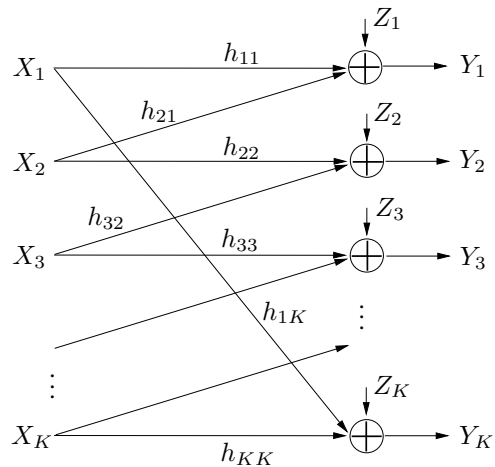}
\caption{$K$-user cyclic Gaussian interference channel} \label{cyclic_ic}
\end{figure}

The $K$-user cyclic Gaussian interference channel (as depicted in Fig.~\ref{cyclic_ic}) has $K$ transmitter-receiver pairs. Each
transmitter tries to communicate with its intended receiver while
causing interference to only one neighboring receiver. Each receiver
receives a signal intended for it and an interference signal from only
one
neighboring sender plus an additive white Gaussian noise (AWGN). As
shown in Fig.~\ref{cyclic_ic}, $X_1, X_2, \cdots X_K$ and $Y_1, Y_2,
\cdots Y_K$ are the complex-valued input and output signals, respectively, and $Z_i
\thicksim \mathcal{CN} (0,\sigma^2)$ is the independent and identically distributed (i.i.d) Gaussian noise at receiver $i$. The input-output model can be written as
\begin{eqnarray}
Y_1 &=& h_{1,1}X_1 + h_{2,1}X_2 + Z_1, \nonumber \\
Y_2 &=& h_{2,2}X_2 + h_{3,2}X_3 + Z_2, \nonumber \\
& \vdots  & \nonumber \\
Y_K &=& h_{K,K}X_K + h_{1,K}X_{1} + Z_K, \nonumber
\label{input_output}
\end{eqnarray}
where each $X_i$ has a power constraint $P_i$ associated with it,
i.e., $\mathbb{E}\left[|x_i|^2 \right] \le P_i$. Here, $h_{i,j}$ is the
channel gain from transmitter $i$ to receiver $j$.

Define the signal-to-noise and
interference-to-noise ratios for each user as follows:
\begin{equation} \label{snr_inr_def}
\mathsf{SNR}_i = \frac{|h_{i,i}|^2P_i}{\sigma^2}  \quad
\mathsf{INR}_i = \frac{|h_{i,i-1}|^2P_{i}}{\sigma^2}, \quad i=1, 2, \cdots, K.
\end{equation}
The $K$-user cyclic Gaussian interference channel is said to be in the
{\it{weak}} interference regime if
\begin{equation} \label{weak_regime_def}
\mathsf{INR}_i \le \mathsf{SNR}_i, \quad \forall i=1, 2, \cdots, K.
\end{equation}
and the {\it{strong}} interference regime if
\begin{equation} \label{strong_regime_def}
\mathsf{INR}_i \ge \mathsf{SNR}_i, \quad \forall i=1, 2, \cdots, K.
\end{equation}
Otherwise, it is said to be in the {\it{mixed}} interference regime.
Throughout this paper, modulo arithmetic is implicitly used on the
user indices, e.g., $K+1=1$ and $1-1=K$. Note that when $K=2$, the
cyclic channel reduces to the conventional two-user interference
channel.

\section{Within Two Bits of the Capacity Region in the Weak Interference Regime}

The generalization of Etkin, Tse and Wang's result to the capacity
region of a general (nonsymmetric) $K$-user cyclic Gaussian
interference channel is significantly more complicated. In the
two-user case, the shape of the Han-Kobayashi achievable rate region
is the union of polyhedrons (each corresponding to a fixed input
distribution) with boundaries defined by rate constraints on $R_1$,
$R_2$, $R_1+R_2$, $2R_1+R_2$ and $2R_2 + R_1$, respectively.
In the multiuser case, to extend Etkin, Tse and Wang's result, one needs
to find a similar rate region characterization for
the general $K$-user cyclic interference channel first.

A key feature of the cyclic Gaussian interference channel model is
that each transmitter sends signal to its intended receiver while
causing interference to {\em only one} of its neighboring receivers;
meanwhile, each receiver receives the intended signal plus the
interfering signal from {\em only one} of its neighboring transmitters.
Using this fact and with the help of Fourier-Motzkin elimination
algorithm, this section shows that the achievable rate region of the
$K$-user cyclic Gaussian interference channel is the union of
polyhedrons with boundaries defined by rate constraints on the
individual rates $R_i$, the sum rate $R_{sum}$, the sum rate plus an
individual rate $R_{sum} + R_i$ ($i=1,2,\cdots, K$), and the sum rate
for arbitrary $l$ adjacent users ($2 \le l < K$). This last rate
constraint on arbitrary $l$ adjacent users' rates is new as compared
with the two-user case.

The preceding characterization together with outer bounds to be proved
later in the section allows us to prove that the capacity region of the
$K$-user cyclic Gaussian interference channel can be achieved to
within a constant gap using the ETW power-splitting strategy in the
weak interference regime.  However, instead of the one-bit result for
the two-user interference channel, this section shows that one can
achieve to within $1\frac{1}{2}$ bits of the capacity region when $K=3$
(with time-sharing), and within two bits of the capacity region for
general $K$. Again, the strong interference regime is treated later.

\subsection{Achievable Rate Region}
\begin{theo} \label{achievable_theo}
Let $\mathcal{P}$ denote the set of probability distributions $P(\cdot)$ that factor as
\begin{eqnarray}
\lefteqn{P(q, w_1,x_1, w_2, x_2, \cdots, w_K, x_K)} \nonumber \\
 && = p(q)p(x_1 w_1|q)p(x_2 w_2|q) \cdots p(x_K w_K|q).
\end{eqnarray}
For a fixed $P \in \mathcal{P}$, let $\mathcal{R}_{\mathrm{HK}}^{(K)}(P)$ be the set of all rate tuples $(R_1, R_2, \cdots, R_K)$ satisfying
\begin{eqnarray}
0 \le R_i &\le& \min\{d_i, a_i + e_{i-1} \}, \label{achievable_Ri} \\
\sum_{j=m}^{m+l-1}R_j &\le& \min \left\{ g_m + \sum_{j=m+1}^{m+l-2}e_j + a_{m+l-1},  \right. \nonumber \\
&& \left.  \qquad \quad \sum_{j=m-1}^{m+l-2}e_j + a_{m+l-1}\right\}, \label{achievable_Rml} \\
 \sum_{j=1}^{K}R_j &\le& \min \left\{ \sum_{j=1}^{K}e_j, r_1, r_2, \cdots, r_K \right\}, \label{achievable_Rsum} \\
\sum_{j=1}^{K}R_j + R_i &\le& a_i + g_i +\sum_{j=1, j \neq i }^{K} e_j ,\label{achievable_Rsi}
\end{eqnarray}
where $a_i, d_i, e_i, g_i$ and $r_i$ are defined as follows:
\begin{eqnarray}
a_i & = & I(Y_i; X_i|W_i,W_{i+1}, Q) \\
d_i & = & I(Y_i; X_i| W_{i+1}, Q) \\
e_i & = & I(Y_i; W_{i+1}, X_i|W_i, Q) \\
g_i & = & I(Y_i; W_{i+1}, X_i| Q)
\end{eqnarray}
\begin{equation}
r_i = a_{i-1} + g_i +\sum_{j=1, j\notin\{i,i-1\}}^{K} e_j,
\end{equation}
and the range of indices are $i, m = 1, 2, \cdots, K$ in
(\ref{achievable_Ri}) and (\ref{achievable_Rsi}), $l = 2, 3, \cdots,
K-1$ in (\ref{achievable_Rml}).
Define
\begin{equation}
\mathcal{R}_{\mathrm{HK}}^{(K)} = \bigcup_{P \in \mathcal{P}} \mathcal{R}_{\mathrm{HK}}^{(K)}(P).
\end{equation}
Then $\mathcal{R}_{\mathrm{HK}}^{(K)}$ is an achievable rate region for the $K$-user cyclic interference channel \footnote{The same achievable rate region has been found independently in \cite{cyclic_first}.}.

\end{theo}

\begin{IEEEproof}
The achievable rate region can be proved by the Fourier-Motzkin
algorithm together with an induction step. The proof follows the
Kobayashi and Han's strategy \cite{HK2007}
of eliminating a common message at each step.  
The details are presented in Appendix \ref{proof_achievable_theo}.
\end{IEEEproof}

In the above achievable rate region, (\ref{achievable_Ri}) is the
constraint on the achievable rate of an individual user,
(\ref{achievable_Rml}) is the constraint on the achievable sum rate
for any $l$ adjacent users ($2 \le l < K$), (\ref{achievable_Rsum}) is
the constraint on the achievable sum rate of all $K$ users, and
(\ref{achievable_Rsi}) is the constraint on the achievable sum rate
for all $K$ users plus a repeated one. We can also think of
(\ref{achievable_Ri})-(\ref{achievable_Rsi}) as the sum-rate
constraints for arbitrary $l$ adjacent users, where $l=1$ for
(\ref{achievable_Ri}), $2 \le l <K$ for (\ref{achievable_Rml}), $l=K$
for (\ref{achievable_Rsum}) and $l=K+1$ for (\ref{achievable_Rsi}).

From (\ref{achievable_Ri}) to (\ref{achievable_Rsi}), there are a
total of $K + K(K-2) + 1 + K =K^2 +1$ constraints. Together they
describe the shape of the achievable rate region under a fixed input
distribution.  The quadratic growth in the number of constraints as a
function of $K$ makes the Fourier-Motzkin elimination of the
Han-Kobayashi region quite complex. The proof in Appendix~\ref{proof_achievable_theo} uses
induction to deal with the large number of the constraints.

As an example, for the two-user Gaussian interference channel, there
are $2^2+1 = 5$ rate constraints, corresponding to that of $R_1$,
$R_2$, $R_1 +R_2$, $2R_1 +R_2$ and $2R_2 +R_1$, as in \cite{HK1981,
HK2007, Chong2006, Tse2007}. Specifically, substituting $K=2$ in
Theorem~\ref{achievable_theo} gives us the following achievable rate
region:
\begin{eqnarray}
0 \le R_1 &\le& \min \{ d_1, a_1 + e_2\},  \label{R1_constraint}\\
0 \le R_2 &\le& \min \{ d_2, a_2 + e_1\},  \label{R2_constraint}\\
R_{1}+R_{2} &\le& \min\{e_1 +e_2, a_1 + g_2, a_2 + g_1  \}, \\
2R_1 + R_2 &\le& a_1 + g_1 + e_2, \\
2R_2 + R_1 &\le& a_2 + g_2 + e_1. \label{R2_lastone}
\end{eqnarray}
The above region for the two-user Gaussian
interference channel is exactly that of Theorem~D in \cite{HK2007}.

\subsection{Capacity Region Outer Bound}

\begin{theo} \label{outerbound_theo}
For the $K$-user cyclic  Gaussian interference channel in the weak
interference regime, the capacity region is included in the set of rate tuples $(R_1, R_2, \cdots, R_K)$ such that
\begin{eqnarray}
R_i &\le& \lambda_i  \label{outer_Ri}, \\
\sum_{j=m}^{m+l-1}R_j &\le& \min \left\{ \gamma_m + \sum_{j=m+1}^{m+l-2}\alpha_j + \beta_{m+l-1},  \right. \nonumber \\
&& \left. \qquad \mu_m + \sum_{j=m}^{m+l-2}\alpha_j + \beta_{m+l-1}\right\}, \label{outer_Rml}\\
\sum_{j=1}^{K}R_j &\le& \min \left\{ \sum_{j=1}^{K}\alpha_j, \rho_1, \rho_2, \cdots, \rho_K \right\}, \label{outer_Rsum} \\
\sum_{j=1}^{K}R_j + R_i &\le& \beta_i + \gamma_i +\sum_{j=1, j \neq i }^{K} \alpha_j, \label{outer_Rsi}
\end{eqnarray}
where the ranges of the indices $i$, $m$, $l$ are as defined in
Theorem~\ref{achievable_theo}, and
\begin{eqnarray}
\alpha_i &=& \log \left(1+ \mathsf{INR}_{i+1} + \frac{\mathsf{SNR}_i}{1+\mathsf{INR}_i} \right) \\
\beta_i &=& \log \left( \frac{1+\mathsf{SNR}_i}{1+\mathsf{INR}_i} \right) \\
\gamma_i &=& \log \left( 1 + \mathsf{INR}_{i+1} + \mathsf{SNR}_i \right) \\
\lambda_i &=& \log (1 + \mathsf{SNR}_i) \\
\mu_i &=& \log (1+ \mathsf{INR}_i) \\
\rho_i &=&\beta_{i-1} + \gamma_i + \sum_{j=1, j\notin\{i,i-1\}}^{K} \alpha_j .
\end{eqnarray}
\begin{IEEEproof}
See Appendix~\ref{appendix_outerbound}.
\end{IEEEproof}

\end{theo}

\subsection{Capacity Region to Within Two Bits}

\begin{theo} \label{twobit_theo}
For the $K$-user cyclic Gaussian interference channel in the weak
interference regime, the fixed ETW power-splitting strategy
achieves to within two bits of the capacity
region\footnote{This paper follows the definition from \cite{Tse2007}
that if a rate tuple $(R_1, R_2, \cdots, R_K)$ is achievable and
$(R_1+b, R_2+b, \cdots, R_K+b)$ is outside the capacity region, then
$(R_1, R_2, \cdots, R_K)$ is within $b$ bits of the capacity region.}.
\end{theo}
\begin{IEEEproof}
Applying the ETW power-splitting strategy (i.e., $\mathsf{INR}_{ip} =
\min (\mathsf{INR}_i, 1)$) to Theorem~\ref{achievable_theo},
parameters $a_i, d_i, e_i, g_i$ can be easily calculated as follows:
\begin{eqnarray}
a_i &=&  \log \left( 2 + \mathsf{SNR}_{ip}  \right) - 1, \\
d_i &=&  \log \left( 2 + \mathsf{SNR}_i  \right) - 1, \label{d_i_2} \\
e_i &=&  \log \left( 1 + \mathsf{INR}_{i+1} + \mathsf{SNR}_{ip} \right) - 1, \\
g_i &=& \log \left( 1 + \mathsf{INR}_{i+1} + \mathsf{SNR}_i \right) - 1,
\end{eqnarray}
where $\mathsf{SNR}_{ip} = |h_{i,i}|^2 P_{ip}/\sigma^2$.
To prove that the achievable rate region in Theorem~\ref{achievable_theo}
with the above $a_i, d_i, e_i, g_i$ is within two bits of the outer
bound in Theorem~\ref{outerbound_theo}, we show that each of the rate
constraints in (\ref{achievable_Ri})-(\ref{achievable_Rsi}) is within
two bits of their corresponding outer bound in
(\ref{outer_Ri})-(\ref{outer_Rsi}) in the weak interference regime,
i.e., the following inequalities hold for all $i$, $m$, $l$ in the
ranges defined in Theorem~\ref{achievable_theo}:
\begin{eqnarray}
\delta_{R_i} & \le& 2, \label{delta_Ri} \\
\delta_{R_m+\cdots+R_{m+l-1}} &\le& 2l, \\
\delta_{R_{sum}} &\le& 2K, \\
\delta_{R_{sum}+R_i} &\le& 2(K+1), \label{delta_RsumRi}
\end{eqnarray}
where $\delta_{(\cdot)}$ is the difference between the achievable rate
in Theorem~\ref{achievable_theo} and its corresponding outer bound in
Theorem ~\ref{outerbound_theo}.  The proof makes use of a set of
inequalities provided in Appendix~\ref{useful_ineq}.

For $\delta_{R_i}$, we have
\begin{eqnarray}
\delta_{R_i} &=& \lambda_i - \min\{d_i, a_i+e_{i-1}\} \nonumber \\
&=& \max \{ \lambda_i - d_i, \lambda_i -(a_i +e_{i-1})\}  \nonumber \\
&\le& 2.
\end{eqnarray}

For $\delta_{R_m + \cdots + R_{m+l-1}}$,  compare the first terms of (\ref{achievable_Rml}) and (\ref{outer_Rml}):
\begin{eqnarray}
\delta_1 &=&  \gamma_m + \sum_{j=m+1}^{m+l-2}\alpha_j + \beta_{m+l-1}  - g_m + \sum_{j=m+1}^{m+l-2}e_j  \nonumber \\
&& + a_{m+l-1}  \nonumber \\
&=&(\gamma_m - g_m) + \sum_{j=m+1}^{m+l-2}(\alpha_j - e_j)  + (\beta_{m+l-1} - a_{m+l-1}) \nonumber \\
&\le& l.
\end{eqnarray}
Similarly, the difference between
the second term of (\ref{achievable_Rml}) and (\ref{outer_Rml}) is bounded by
\begin{eqnarray}
\delta_2 &=& \mu_m + \sum_{j=m}^{m+l-2}\alpha_j + \beta_{m+l-1} - \sum_{j=m-1}^{m+l-2}e_j + a_{m+l-1} \nonumber \\
&=& (\mu_m - e_{m-1}) + \sum_{j=m}^{m+l-2}(\alpha_j - e_j)  \nonumber \\
&& + ( \beta_{m+l-1} - a_{m+l-1}) \nonumber \\
&\le& l + 1.
\end{eqnarray}
Finally, applying the fact that
\begin{equation}
\min\{x_1, y_1\} - \min \{x_2, y_2 \} \le \max \{x_1-x_2, y_1-y_2\}, \nonumber
\end{equation}
we obtain
\begin{equation}
\delta_{R_m+ \cdots + R_{m+l-1}} \le \max \{\delta_1, \delta_2 \} \le l+1.
\end{equation}

For $\delta_{R_{sum}}$, the difference between the first terms of (\ref{achievable_Rsum}) and (\ref{outer_Rsum}) is bounded by
\begin{eqnarray}
\sum_{j=1}^{K}\alpha_j - \sum_{j=1}^{K}e_j =\sum_{j=1}^{K}(\alpha_j - e_j) \le K.
\end{eqnarray}
In addition,
\begin{eqnarray}
\rho_i - r_i &=& \beta_{i-1} + \gamma_i + \sum_{j=1, j\notin\{i,i-1\}}^{K} \alpha_j  \nonumber \\
&& -  a_{i-1} + g_i +\sum_{j=1, j\notin\{i,i-1\}}^{K} e_j   \nonumber \\
&=& (\beta_{i-1} - a_{i-1}) + (\gamma_i - g_i)  \nonumber \\
&&+ \sum_{j=1, j\notin\{i,i-1\}}^{K} (\alpha_j - e_j) \nonumber \\
&\le& K
\end{eqnarray}
for $i=1,2,\cdots, K$.
As a result, the gap on the sum rate is bounded by
\begin{eqnarray}
\delta_{R_{sum}} &=& \min \left\{ \sum_{j=1}^{K}\alpha_j, \rho_1, \rho_2, \cdots, \rho_K \right\}  \nonumber \\
&&- \min \left\{ \sum_{j=1}^{K}e_j, r_1, r_2, \cdots, r_K \right\} \nonumber \\
&\le& \max \left\{ \sum_{j=1}^{K}(\alpha_j-e_j), \rho_1 - r_1,  \right. \nonumber \\
&& \left. \rho_2-r_2, \cdots,  \rho_K - r_K \right \} \nonumber \\
&\le& K.
\end{eqnarray}

For $R_{sum} + R_i$, we have
\begin{eqnarray}
\delta_{R_{sum}+R_i} &=& \beta_i + \gamma_i +\sum_{j=1, j \neq i }^{K} \alpha_j  - a_i + g_i + \sum_{j=1, j \neq i }^{K} e_j  \nonumber \\
&=& (\beta_i-a_i) + (\gamma_i - g_i) + \sum_{j=1, j \neq i }^{K} (\alpha_j - e_j) \nonumber \\
&\le& K+1
\end{eqnarray}

Since the inequalities in (\ref{delta_Ri})-(\ref{delta_RsumRi})
hold for all the ranges of $i$, $m$, and $l$ defined in
Theorem~\ref{achievable_theo}, this proves that the ETW power-splitting strategy achieves to within two bits of the capacity region in the weak
interference regime.
\end{IEEEproof}

\subsection{3-User Cyclic Gaussian Interference Channel Capacity
Region to Within $1\frac{1}{2}$ Bits}

Chong, Motani and Garg
\cite{Chong2006} showed that by time-sharing with marginalized
versions of the input distribution, the Han-Kobayashi region for the
two-user interference channel as stated in
(\ref{R1_constraint})-(\ref{R2_lastone}) can be further simplified by
removing the $a_1 + e_2$ and $a_2 + e_1$ terms from
(\ref{R1_constraint}) and (\ref{R2_constraint}) respectively. The
resulting rate region without these two terms is
proved to be equivalent to the original Han-Kobayashi region
(\ref{R1_constraint})-(\ref{R2_lastone}).

This section shows that the aforementioned time-sharing technique can
be applied to the $3$-user cyclic interference channel (but
not to $K \ge 4$). By a similar time-sharing strategy, the second rate
constraint on $R_1, R_2$ and $R_3$ can be removed, and the achievable
rate region can be shown to be within $1\frac{1}{2}$ bits of the
capacity region in the weak interference regime.


\begin{theo} \label{3_user_ic}
Let $\mathcal{P}_3$ denote the set of probability distributions $P_3(\cdot)$ that factor as
\begin{eqnarray}
\lefteqn{P_3(q, w_1, x_1, w_2, x_2, w_3, x_3) }\nonumber \\
&=& p(q)p(x_1 w_1|q)p(x_2 w_2|q)p(x_3 w_3|q).
\end{eqnarray}
For a fixed $P_3 \in \mathcal{P}_3$, let $\mathcal{R}_{\textrm{HK-TS}}^{(3)}(P_3)$ be the set of all rate tuples $(R_1, R_2, R_3)$ satisfying
\begin{eqnarray}
R_i &\le& d_i, \quad i=1, 2, 3, \label{R_new_R}\\
R_1 + R_2 &\le& \min\{ g_1 +a_2, e_3 + e_1 + a_2 \}, \label{R_new_R1+R2} \\
R_2 + R_3 &\le& \min\{ g_2 +a_3, e_1 + e_2 + a_3 \}, \label{R_new_R2+R3}\\
R_3 + R_1 &\le& \min\{ g_3 +a_1, e_2 + e_3 + a_1 \},  \label{R_new_R3+R1} \\
R_1 + R_2 + R_3 &\le& \min \{e_1+e_2+e_3, a_3 + g_1 +e_2,  \nonumber \\
&& a_1 + g_2 + e_3, a_2 + g_3 +e_1\}, \label{R_new_R1+R2+R3} \\
2R_1 + R_2 +R_3 &\le& a_1 + g_1 + e_2 + e_3, \label{R_new_Rsum+R1} \\
R_1 + 2R_2 +R_3 &\le& a_2 + g_2 + e_3 + e_1, \label{R_new_Rsum+R2}\\
R_1 + R_2 +2R_3 &\le& a_3 + g_3 + e_1 + e_2, \label{R_new_Rsum+R3}
\end{eqnarray}
where $a_i, d_i, e_i, g_i$ are as defined before. Define
\begin{equation}
\mathcal{R}_{\textrm{HK-TS}}^{(3)} = \bigcup_{P_3 \in \mathcal{P}_3}
\mathcal{R}_{\textrm{HK-TS}}^{(3)}(P_3).
\end{equation}
Then, $\mathcal{R}_{\textrm{HK-TS}}^{(3)}$ is an achievable rate region for the $3$-user cyclic Gaussian interference channel.
Further, when $P_3$ is set according to the ETW power-splitting strategy, the
rate region $R_{\textrm{HK-TS}}^{(3)}(P_3)$ is within $1\frac{1}{2}$
bits of the capacity region in the weak interference regime.
\end{theo}

\begin{IEEEproof}
We prove the achievability of $\mathcal{R}_{\textrm{HK-TS}}^{(3)}$ by
showing that $\mathcal{R}_{\textrm{HK-TS}}^{(3)}$ is equivalent to
$\mathcal{R}_{\textrm{HK}}^{(3)}$.  First, since
$\mathcal{R}_{\textrm{HK}}^{(3)}$ contains an extra constraint on each of
$R_1, R_2$ and $R_3$ (see (\ref{achievable_Ri})), it immediately
follows that
\begin{equation}
\mathcal{R}_{\textrm{HK}}^{(3)} \subseteq \mathcal{R}_{\textrm{HK-TS}}^{(3)}.
\end{equation}
In Appendix~\ref{R_NEW_R_HK_inclusion}, it is shown that the
inclusion also holds the other way around. Therefore,
$\mathcal{R}_{\textrm{HK}}^{(3)} = \mathcal{R}_{\textrm{HK-TS}}^{(3)}$
and as a result, $\mathcal{R}_{\textrm{HK-TS}}^{(3)}$ is achievable.

Applying the ETW power-splitting strategy (i.e., $\mathsf{INR}_{ip} =
\min\{\mathsf{INR}_i, 1 \}$ and $Q$ is fixed) to
$\mathcal{R}^{(3)}_{\textrm{HK-TS}}(P_3)$, and following along the same
line of the proof of Theorem~\ref{twobit_theo}, we obtain
\begin{eqnarray}
\delta_{R_i} &\le& 1, \\
\delta_{R_i+R_{i+1}} &\le& 3, \\
\delta_{R_{sum}} &\le& 3, \\
\delta_{R_{sum} + R_i} &\le& 4,
\end{eqnarray}
where $i=1, 2, 3$. It then follows that the gap to the capacity region
is at most $1\frac{1}{2}$ bits in the weak interference regime.
\end{IEEEproof}

As shown in Appendix~\ref{R_NEW_R_HK_inclusion}, the rate region
(\ref{R_new_R})-(\ref{R_new_Rsum+R3}) is obtained by taking the union
over the achievable rate regions with input distributions $P_3, P_3^{*}, P_3^{**}$ and $P_3^{***}$, where $P_3^{*},
P_3^{**}$ and $P_3^{***}$ are the marginalized versions of $P_3$.
Thus, to achieve within $1\frac{1}{2}$ bits of the capacity region,
one needs to time-share among the ETW power-splitting and its three
marginalized variations, rather than using the fixed ETW's input alone.

The key improvement of $\mathcal{R}_{\textrm{HK-TS}}^{(3)}$ over
$\mathcal{R}_{\textrm{HK}}^{(3)}$ is the removal of term $a_i +
e_{i-1}$ in (\ref{achievable_Ri}) using a time-sharing technique.
However, the results in Appendix~\ref{R_NEW_R_HK_inclusion} hold only
for $K=3$. When $K \ge 4$, it is easy to verify that
$\mathcal{R}_{\textrm{HK-TS}}^{(4)}(P_4)$ is not within the union of
$\mathcal{R}_{\textrm{HK}}^{(4)}(P_4)$ and its marginalized
variations, i.e., $\mathcal{R}_{\textrm{HK}}^{(4)} \nsubseteq
\mathcal{R}_{\textrm{HK-TS}}^{(4)}$. Therefore, the techniques used in
this paper only allow the two-bit result to be sharpened to
a $1\frac{1}{2}$-bit result for the three-user cyclic Gaussian
interference channel, but not for $K \ge 4$.

\section{Capacity Region in the Strong Interference Regime}
The results so far pertain only to the weak interference regime, where $\mathsf{SNR}_i \ge \mathsf{INR}_i$, $\forall i$. In the strong interference regime, where $\mathsf{SNR}_i \le \mathsf{INR}_i$, $\forall i$, the capacity result in \cite{HK1981} \cite{Sato} for the two-user Gaussian interference channel can be easily extended to the $K$-user  cyclic case.

\begin{theo} \label{strong_theo}
For the $K$-user cyclic Gaussian interference channel in the strong
interference regime, the capacity region is given by the set of
$(R_1,R_2,\cdots,R_K)$ such that \footnote{This capacity result was also recently obtained in \cite{multiuser_cyclic_ic}.}
\begin{eqnarray} \label{capacity_strong}
\left\{
  \begin{array}{l}
R_i \le \log (1 + \mathsf{SNR}_i) \\
R_i + R_{i+1} \le \log (1 + \mathsf{SNR}_i + \mathsf{INR}_{i+1}),
  \end{array}
\right.
\end{eqnarray}
for $i=1, 2, \cdots, K$. In the very strong interference
regime where $\mathsf{INR}_i \ge (1 +  \mathsf{SNR}_{i-1})\mathsf{SNR}_i,
\forall i$, the capacity region is the set of $(R_1,R_2,\cdots,R_K)$ with
\begin{equation} \label{capacity_very_strong}
R_i \le \log (1 + \mathsf{SNR}_i), \;\; i=1, 2, \cdots, K.
\end{equation}
\end{theo}

\begin{IEEEproof}
{\em{Achievability}}: It is easy to see that (\ref{capacity_strong})
is in fact the intersection of the capacity regions of
$K$ multiple-access channels:
\begin{equation}
\label{capacityregion_mac}
\bigcap_{i=1}^{K}\left\{
(R_i,R_{i+1}) \left|  \begin{array}{l}
R_i \le \log (1 + \mathsf{SNR}_i) \\
R_{i+1} \le \log (1 + \mathsf{INR}_{i+1}) \\
R_i + R_{i+1} \le \log (1 + \mathsf{SNR}_i + \mathsf{INR}_{i+1}).
  \end{array}
\right.  \right \}.
\end{equation}
Each of these regions corresponds to that of a multiple-access channel with $W_i^n$ and $W_{i+1}^n$ as inputs and $Y_i^n$ as output (with $U_i^n=U_{i+1}^n=\emptyset$).  Therefore, the rate region (\ref{capacity_strong}) can be achieved by setting  all the input signals to be common messages.  This completes the achievability part.

{\em{Converse}}: The converse proof follows the idea of \cite{Sato}. The key ingredient is to show that for a genie-aided Gaussian interference channel to be defined later, in the strong interference regime, whenever a rate tuple $(R_1, R_2, \cdots, R_K)$ is achievable, i.e., $X_i^n$ is decodable at receiver $i$, $X_i^n$ must also be decodable at $Y_{i-1}^n$,  $i=1,2,\cdots, K$.

The genie-aided Gaussian interference channel is defined by the Gaussian interference channel (see Fig.~\ref{cyclic_ic}) with genie $X_{i+2}^n$ given to receiver $i$. The capacity region of the $K$-user cyclic Gaussian interference channel must reside inside the capacity region of the genie-aided one.


Assume that a rate tuple $(R_1, R_2, \cdots, R_K)$ is achievable for the $K$-user cyclic Gaussian interference channel. In this case, after $X_i^n$ is decoded, with the knowledge of the genie $X_{i+2}^n$, receiver $i$ can construct the following signal:
\begin{eqnarray}
\widetilde{Y}_i^n &=& \frac{h_{i+1,i+1}}{h_{i+1,i}}(Y_i^n - h_{i,i}X_i^n) + h_{i+2, i+1}X_{i+2}^n \nonumber \\
&=& h_{i+1, i+1}X_{i+1}^n + h_{i+2, i+1}X_{i+2}^n + \frac{h_{i+1,i+1}}{h_{i+1,i}}Z_{i}^n, \nonumber
\end{eqnarray}
which contains the signal component of $Y_{i+1}^n$ but with less noise since $|h_{i+1,i}| \ge |h_{i+1,i+1}|$ in the strong interference regime. Now, since $X_{i+1}^n$ is decodable at receiver $i+1$, it must also be decodable at receiver $i$ using the constructed $\widetilde{Y}_i^n$. Therefore, $X_i^n$ and
$X_{i+1}^n$ are both decodable at receiver $i$. As a result, the achievable rate region of $(R_i, R_{i+1})$ is bounded by the capacity region of the multiple-access channel $(X_i^n, X_{i+1}^n, Y_i^n)$, which is shown in (\ref{capacityregion_mac}). Since (\ref{capacityregion_mac}) reduces to (\ref{capacity_strong}) in the strong interference regime, we have shown that (\ref{capacity_strong}) is an outer bound of the $K$-user
cyclic Gaussian interference channel in the strong  interference regime. This completes the converse proof.

In the very strong interference regime where $\mathsf{INR}_i \ge (1 +  \mathsf{SNR}_{i-1})\mathsf{SNR}_i, \forall i$, it is easy to verify that the second constraint in (\ref{capacity_strong}) is no longer active. This results in the capacity region (\ref{capacity_very_strong}).
\end{IEEEproof}

\section{Symmetric Channel and Generalized Degrees of Freedom}

Consider the symmetric cyclic Gaussian interference channel, where all
the direct links from the transmitters to the receivers share the same channel
gain and all the cross links share another same channel gain. In addition, all the input signals have the same power
constraint $P$, i.e., $\mathbb{E}\left[|X_i|^2\right] \le P, \forall i$.

The symmetric capacity of the $K$-user interference channel is defined as
\begin{eqnarray} \label{C_sym}
C_{sym} = \left\{
  \begin{array}{l}
\textrm{maximize \ min}\{R_1, R_2, \cdots, R_K\} \\
\textrm{subject to \ } \; (R_1, R_2, \cdots, R_K) \in \mathcal{R}
  \end{array}
\right.
\end{eqnarray}
where $\mathcal{R}$ is the capacity region of the $K$-user
interference channel. For the symmetric interference channel, $C_{sym}
= \frac{1}{K}C_{sum}$, where $C_{sum}$ is the sum capacity. As a direct consequence of Theorem~\ref{twobit_theo} and
Theorem~\ref{strong_theo}, the generalized degree of freedom
of the symmetric capacity for the symmetric cyclic channel can be derived as follows.

\begin{corol} \label{dymmetric_gdof}
For the $K$-user symmetric cyclic Gaussian interference channel,
\begin{eqnarray} \label{C_sym}
d_{sym} = \left\{
  \begin{array}{l}
\min \left\{ \max \{\alpha, 1-\alpha \}, 1 -\frac{\alpha}{2} \right\}, \;\; 0 \le \alpha < 1 \\
\min \{\frac{\alpha}{2}, 1 \}, \qquad  \qquad  \qquad \qquad \qquad \alpha \; \ge 1
  \end{array}
\right.
\end{eqnarray}
where $d_{sym}$ is the generalized degrees of freedom of the symmetric capacity.
\end{corol}

Note that the above $d_{sym}$ for the $K$-user cyclic interference channel with symmetric channel parameters is the same as that of the two-user interference channel derived in \cite{Tse2007}.

\section{Conclusion}

This paper investigates the capacity and the coding strategy for the
$K$-user cyclic Gaussian interference channel. Specifically, this paper shows that in the weak interference
regime, the ETW power-splitting strategy achieves to within two bits
of the capacity region.  Further, in the special case of $K=3$ and
with the help of a time-sharing technique, one can achieve to within
$1\frac{1}{2}$ bits of the capacity region in the weak interference
regime.

The capacity result for the $K$-user cyclic Gaussian interference
channel in the strong interference regime is a straightforward
extension of the corresponding two-user case. However, in the mixed
interference regime, although the constant gap result may well
continue to hold, the proof becomes considerably more complicated, as
different mixed scenarios need to be enumerated and the
corresponding outer bounds derived.

\appendix

\subsection{Proof of Theorem~\ref{achievable_theo}}
\label{proof_achievable_theo}

For the two-user interference channel, Kobayashi and Han \cite{HK2007}
gave a detailed Fourier-Motzkin elimination procedure for the
achievable rate region. The Fourier-Motzkin elimination for the
$K$-user cyclic interference channel involves $K$ elimination steps.
The complexity of the process increases with each step.  Instead of
manually writing down all the inequalities step by step, this appendix
uses mathematical induction to derive the final result.

This achievability proof is based on the application of coding scheme
in \cite{Chong2006} (also referred as the multi-level coding in
\cite{Viswanath_compound}) to the multi-user setting. Instead of using
the original code construction of \cite{HK1981}, the following strategy is used in which each common
message $W_i, i=1,2,\cdots,K$ serves to generate $2^{nT_i}$ cloud
centers $W_i(j), j=1,2,\cdots, 2^{nT_i}$, each of which is surrounded
by $2^{nS_i}$ codewords $X_i(j, k), k=1,2,\cdots,2^{nS_i}$.  This
results in achievable rate region expressions expressed in terms of
$(W_i,X_i,Y_i)$ instead of $(U_i,W_i,Y_i)$.
For the two-user interference channel, Chong, Motani and Garg
\cite[Lemma 3]{Chong2006} made a further simplification to the
achievalbe rate region expression. They observed that in the
Han-Kobayashi scheme, the common message $W_i$ is only required to be
correctly decoded at the intended receiver $Y_i$ and an incorrectly
decoded $W_i$ at receiver $Y_{i-1}$ does not cause an error event.
Based on this observation, they concluded that for the multiple-access channel with input $(U_i,W_i,W_{i+1})$ and output $Y_i$,
the rate constraints on
common messages $T_i$, $T_{i+1}$ and $T_i + T_{i+1}$
are in fact irrelevant to the decoding error
probabilities and can be removed, i.e., the rates $(S_i,T_i,T_{i+1})$
are constrained by only the following set of inequalities:
\begin{eqnarray}
\label{cmg_begin}
S_i & \le & a_i = I(Y_i; X_i|W_i,W_{i+1}, Q) \\
S_i + T_i & \le & d_i = I (Y_i; X_i| W_{i+1}, Q) \\
S_i + T_{i+1} & \le & e_i = I (Y_i; W_{i+1}, X_i|W_i, Q) \\
S_i + T_i + T_{i+1} & \le & g_i = I (Y_i; W_{i+1}, X_i| Q) \\
S_i, T_i, T_{i+1} & \ge & 0
\label{cmg_end}
\end{eqnarray}
Now, compare the $K$-user cyclic interference channel with the
two-user interference channel, it is easy to see that in both channel
models, each receiver only sees interference from one neighboring
transmitter. This makes the decoding error probability analysis for both channel
models the same.  Therefore, the set of rates $\mathcal{R}(R_1, R_2,
\cdots, R_K)$, where $R_i=S_i+T_i$, with $(S_i,T_i)$ satisfy
(\ref{cmg_begin})-(\ref{cmg_end}) for $i=1, 2, \cdots, K$,
characterizes an achievable rate region for the $K$-user cyclic
interference channel.

The first step of using the Fourier-Motzkin algorithm is to eliminate
all private messages $S_i$ by substituting $S_i =R_i - T_i$ into the $K$
polymatroids (\ref{cmg_begin})-(\ref{cmg_end}). This results in the
following $K$ polymatroids without $S_i$:
\begin{eqnarray}
R_i - T_i &\le&  a_i,  \label{Polymatroid_no_Si_start}  \\
R_i  &\le&   d_i,  \\
R_i -T_i + T_{i+1} &\le&   e_i,  \\
R_i + T_{i+1} &\le& g_i,  \\
-R_i &\le& 0, \label{Polymatroid_no_Si_end}
\end{eqnarray}
where $i=1,2, \cdots, K$.

Next, use Fourier-Motzkin algorithm to eliminate common message rates
$T_1$, $T_2$, $\cdots$, $T_K$ in a step-by-step process so that after
$n$ steps, common variables $(T_1, \cdots, T_n)$ are eliminated. The
induction hypothesis is the following $5$ different groups of inequalities, which is
assumed to be obtained at the end of the $n$th elimination step:

(a) Inequalities not including private or common variables $S_{i}$ and $T_{i}, i=1, 2, \cdots, K$:
\begin{eqnarray}
R_i &\le& d_i, \quad i=1,2, \cdots, K  \label{grp_1_R_i} \\
-R_i &\le& 0, \quad i=1, 2, \cdots, n  \\
R_K + R_1 &\le& g_{K} + a_{1},  \\
R_{m} &\le& a_m + e_{m-1},  \\
\sum_{j=l}^{m} R_j &\le& \min \left \{ g_l + \sum_{i=l+1}^{m-1}e_j + a_m,  \sum_{j=l-1}^{m-1}e_j + a_m \right \}, \label{reason_of_redundant_3} \nonumber \\
&& \\
\sum_{j=1}^{m} R_j &\le& g_1 + \sum_{j=2}^{m-1}e_j + a_m, \label{reason_of_redundant_4} \\
\sum_{j=K}^{m} R_j &\le& g_K + \sum_{j=1}^{m-1}e_j + a_m,  \label{grp_1_Rkm}
\end{eqnarray}
where $m=2, 3, \cdots, n$ and $l=2, 3, \cdots, m-1$.

(b) Inequalities including $T_{K}$ but not including $T_{n+1}$:
\begin{eqnarray}
R_K - T_K &\le& a_K,  \label{grp_2_RKTK}\\
-R_K -T_K &\le& 0,  \\
-T_K &\le& 0,  \\
\sum_{j=K}^{p}R_j - T_K &\le& \sum_{j=K}^{p-1}e_j + a_p,  \label{grp_2_RkpTk}
\end{eqnarray}
where $p=1,2,\cdots, n$.

(c) All other inequalities not including $T_{n+1}$:
\begin{equation}
R_{n+1} +T_{n+2} \le g_{n+1}, \label{grp_3}
\end{equation}
and all the polymatroids in (\ref{Polymatroid_no_Si_start})-(\ref{Polymatroid_no_Si_end}) indexed from $n+2$ to $K-1$.

(d) Inequalities including $T_{n+1}$ with a plus sign:
\begin{eqnarray}
T_{n+1} &\le& e_n,  \label{grp_4_Tn1}\\
-R_{n+1} + T_{n+1} &\le& 0, \\
\sum_{j=l}^{n}R_j + T_{n+1} &\le& \min \left\{\sum_{j=l-1}^{n}e_j, g_l + \sum_{j=l+1}^{n}e_j  \right \}, \nonumber \\
&& \\
\sum_{j=1}^{n}R_j +T_{n+1} &\le& g_1 + \sum_{j=2}^{n}e_j, \\
\sum_{j=K}^{n}R_j +T_{n+1} &\le&  g_K + \sum_{j=1}^{n}e_j, \label{grp_4_RKnTn1}\\
\sum_{j=K}^{n}R_j + T_{n+1} -T_{K} &\le& \sum_{j=K}^{n}e_j, \label{grp_4_RKnTn1TK}
\end{eqnarray}
where $l$ goes from $2$ to $n$.

(e) Inequalities including $T_{n+1}$ with a minus sign:
\begin{eqnarray}
R_{n+1} - T_{n+1} &\le& a_{n+1}, \label{grp_5_Rn1Tn1}\\
R_{n+1} - T_{n+1} + T_{n+2} &\le& e_{n+1}, \label{grp_5_Rn1Tn1Tn2}\\
-T_{n+1} &\le& 0.  \label{grp_5_minus_Tn1}
\end{eqnarray}


It is easy to verify the correctness of inequalities
(\ref{grp_1_R_i})-(\ref{grp_5_minus_Tn1}) for $n=2$. We next show that
for $n < K-2$, if at the end of step $n$, the inequalities in
(\ref{grp_1_R_i})-(\ref{grp_5_minus_Tn1}) hold,
then they must also hold at the end of step $n+1$.
Towards this end, we follow the Fourier-Motzkin algorithm
\cite{HK2007} by first adding up all the inequalities in
(\ref{grp_4_Tn1})-(\ref{grp_4_RKnTn1TK}) with each of the inequalities in
(\ref{grp_5_Rn1Tn1})-(\ref{grp_5_minus_Tn1}) to eliminate $T_{n+1}$.
This  results in the following three groups of inequalities:

(a) Inequalities due to (\ref{grp_5_Rn1Tn1}):

\begin{eqnarray}
R_{n+1} &\le& a_{n+1} + e_{n}, \label{start} \\
0 &\le&  a_{n+1}, \label{redundant_1} \\
\sum_{j=l}^{n+1}R_j &\le& \min \left\{\sum_{j=l-1}^{n}e_j + a_{n+1}, \right. \nonumber \\
&& \left.  \qquad \quad g_l + \sum_{j=l+1}^{n}e_j + a_{n+1} \right \}, \\
\sum_{j=1}^{n+1}R_j &\le& g_1 + \sum_{j=2}^{n}e_j + a_{n+1}, \\
\sum_{j=K}^{n+1}R_j &\le&  g_K + \sum_{j=1}^{n}e_j + a_{n+1}, \\
\sum_{j=K}^{n+1}R_j - T_{K} &\le& \sum_{j=K}^{n}e_j + a_{n+1},
\end{eqnarray}
where $l=2, 3, \cdots, n$.

(b) Inequalities due to (\ref{grp_5_Rn1Tn1Tn2}):
\begin{eqnarray}
R_{n+1} + T_{n+2} &\le& e_n + e_{n+1}, \\
T_{n+2} &\le& e_{n+1}, \\
\sum_{j=l}^{n+1}R_j + T_{n+2} &\le& \min \left\{\sum_{j=l-1}^{n+1}e_j , g_l + \sum_{j=l+1}^{n+1}e_j \right \}, \nonumber \\
&& \\
\sum_{j=1}^{n+1}R_j + T_{n+2} &\le& g_1 + \sum_{j=2}^{n+1}e_j, \\
\sum_{j=K}^{n+1}R_j + T_{n+2} &\le&  g_K + \sum_{j=1}^{n+1}e_j, \\
\sum_{j=K}^{n+1}R_j + T_{n+2} - T_{K} &\le& \sum_{j=K}^{n+1}e_j,
\end{eqnarray}
where $l=2, 3, \cdots, n$.

(c) Inequalities due to (\ref{grp_5_minus_Tn1}):
\begin{eqnarray}
0 &\le& e_n,  \label{redundant_2}\\
-R_{n+1} &\le& 0, \label{end}\\
\sum_{j=l}^{n}R_j  &\le& \min \left\{\sum_{j=l-1}^{n}e_j, g_l + \sum_{j=l+1}^{n}e_j  \right \}, \label{redundant_3}\\
\sum_{j=1}^{n}R_j  &\le& g_1 + \sum_{j=2}^{n}e_j, \label{redundant_4} \\
\sum_{j=K}^{n}R_j  &\le&  g_K + \sum_{j=1}^{n}e_j, \label{redundant_5} \\
\sum_{j=K}^{n}R_j -T_{K} &\le& \sum_{j=K}^{n}e_j, \label{redundant_6}
\end{eqnarray}
where $l=2, 3, \cdots, n$.

Inspecting the above three groups of inequalities, we can see that
(\ref{redundant_1}) and (\ref{redundant_2}) are obviously redundant.
Also, (\ref{redundant_3}) is redundant due to
(\ref{reason_of_redundant_3}),  (\ref{redundant_4}) is redundant due
to (\ref{reason_of_redundant_4}),  (\ref{redundant_5}) is redundant
due to (\ref{grp_1_Rkm}), and (\ref{redundant_6}) is redundant due to
(\ref{grp_2_RkpTk}). Now, with these six redundant inequalities
removed, the above three groups of inequalities in
(\ref{start})-(\ref{end}) together with
(\ref{grp_1_R_i})-(\ref{grp_3}) form the set of inequalities at the
end of step $n+1$. It can be verified that this new set of inequalities is
exactly (\ref{grp_1_R_i})-(\ref{grp_5_minus_Tn1}) with  $n$ replaced
by $n+1$. This completes the induction part.

Now, we proceed with the $(K-1)$th step. At the end of this step,
$T_1, T_2, \cdots, T_{K-1}$ would all be removed and only $T_K$ would
remain.  Because of the cyclic nature of the channel, the set of
inequalities (\ref{grp_1_R_i})-(\ref{grp_5_minus_Tn1}) needs to be
modified for this $n=K-1$ case. It can be verified that at the end of
the $(K-1)${th} step of Fourier-Motzkin algorithm, we obtain the
following set of inequalities:

(a) Inequalities not including $T_K$: (\ref{grp_1_R_i})-(\ref{grp_1_Rkm}) with $n$ replaced by $K-1$ and
\begin{eqnarray}
\sum_{j=1}^{K}R_j  \le \sum_{j=1}^{K}e_j.   \label{sum_rate_1}
\end{eqnarray}

(b) Inequalities including $T_K$ with a plus sign: (\ref{grp_4_Tn1})-(\ref{grp_4_RKnTn1}) with $n$ replace by $K-1$.
Note that, (\ref{grp_4_RKnTn1TK}) becomes (\ref{sum_rate_1}) when $n=K-1$.

(c) Inequalities including $T_K$ with a minus sign:
\begin{eqnarray}
R_K - T_K &\le& a_K,  \\
\sum_{j=K}^{l}R_j -T_K &\le& \sum_{j=K}^{l-1}e_j + a_l, \\
-T_K &\le& 0,
\end{eqnarray}
where $l=1, 2, \cdots, K-1$.

In the $K$th step (final step) of the Fourier-Motzkin algorithm,
$T_K$ is eliminated by adding each of the inequalities involving $T_K$
with a plus sign and each of the inequalities involving $T_K$ with a
minus sign to obtain new inequalities not involving $T_K$. (This is
quite similar to the procedure of obtaining
(\ref{start})-(\ref{redundant_6}).) Finally,
after removing all the redundant inequalities, we obtain the set of
inequalities in Theorem~\ref{achievable_theo}.

\subsection{Proof of Theorem~\ref{outerbound_theo}}
\label{appendix_outerbound}

We will prove the outer bounds from (\ref{outer_Ri}) to (\ref{outer_Rsi}) one by one.

First, (\ref{outer_Ri}) is simply the cut-set upper bound for user $i$.

Second,  (\ref{outer_Rml}) is the bound on the sum-rate of $l$ adjacent users starting from $m$. According to Fano's inequality, for a block of length $n$, we have
\begin{eqnarray}
\lefteqn{n\left(\sum_{j=m}^{m+l-1}R_j - \epsilon_n \right)}  \nonumber \\
&\le& \sum_{j=m}^{m+l-1}I(x_j^n; y_j^n)\nonumber \\
&\stackrel{(a)}{\le}& h(y_m^n) -h(y_m^n | x_m^n) + \sum_{j=m+1}^{m+l-2}I(x_j^n; y_j^n s_j^n) \nonumber \\
&& + I(x_{m+l-1}^n; y_{m+l-1}^n | x_{m+l}^{n}) \nonumber  \\
&=& h(y_m^n) - h(s_{m+1}^n)   \nonumber \\
&&+ \sum_{j=m+1}^{m+l-2} \left[ h(s_j^n) -h(z_{j-1}^n) + h(y_j^n|s_j^n) - h(s_{j+1}^n) \right]  \nonumber \\
&& + h(h_{m+l-1, m+l-1}x_{m+l-1}^n + z_{m+l-1}^n) - h(z_{m+l-1}^n)  \nonumber \\
&=& h(y_m^n)- h(z_{m+l-1}^n) + \sum_{j=m+1}^{m+l-2}\left[ h(y_j^n|s_j^n) -h(z_{j-1}^n)   \right]  \nonumber \\
&&  +h(h_{m+l-1, m+l-1}x_{m+l-1}^n + z_{m+l-1}^n) \nonumber \\
&& - h(h_{m+l-1, m+l-2}x_{m+l-1}^n + z_{m+l-2}^n)  \nonumber\nonumber \\
&\stackrel{(b)}{\le}& n \left( \gamma_m + \sum_{j=m+1}^{m+l-2}\alpha_j + \beta_{m+l-1}  \right), \label{bound_Rml_1}
\end{eqnarray}
where in (a) we give genie $s_j^n$ to $y_j^n$ for $m+1 \le j \le m+l-2$ and $x_{m+l}^n$ to $y_{m+l-1}^n$ (genies $s_j^n$ are as defined in \cite[Theorem 2]{zhou_kuser}), and (b) comes from the fact \cite{Tse2007} that Gaussian inputs maximize 1) entropy $h(y_m^n)$, 2)  conditional entropy $h(y_j^n|s_j^n)$ for any $j$, and 3) entropy difference $h(h_{m+l-1, m+l-1}x_{m+l-1}^n + z_{m+l-1}^n) - h(h_{m+l-1, m+l-2}x_{m+l-1}^n + z_{m+l-2}^n)$. This proves the first bound in (\ref{outer_Rml}).

Similarly, the second upper bound of (\ref{outer_Rml}) can be obtained by giving genie $s_j^n$ to $y_j^n$ for $m \le j \le m+l-2$ and $x_{m+l}^n$ to $y_{m+l-1}^n$:
\begin{eqnarray}
\lefteqn{n \left(\sum_{j=m}^{m+l-1}R_j - \epsilon_n \right)} \nonumber \\
&\le& \sum_{j=m}^{m+l-1}I(x_j^n; y_j^n)  \nonumber \\
&\le& \sum_{j=m}^{m+l-2}I(x_j^n; y_j^n s_j^n) + I(x_{m+l-1}^n; y_{m+l-1}^n | x_{m+1}^{n})   \nonumber  \\
&=& \sum_{j=m}^{m+l-2} \left[ h(s_j^n) -h(z_{j-1}^n) + h(y_j^n|s_j^n) - h(s_{j+1}^n) \right]   \nonumber \\
&&  + h(h_{m+l-1, m+l-1}x_{m+l-1}^n + z_{m+l-1}^n) - h(z_{m+l-1}^n)  \nonumber \\
&=& h(s_m^n)- h(z_{m+l-1}^n) + \sum_{j=m}^{m+l-2}\left[ h(y_j^n|s_j^n) -h(z_{j-1}^n)   \right]  \nonumber \\
&&  +h(h_{m+l-1, m+l-1}x_{m+l-1}^n + z_{m+l-1}^n) \nonumber \\
&& - h(h_{m+l-1, m+l-2}x_{m+l-1}^n + z_{m+l-2}^n) \nonumber\nonumber \\
&\le& n \left( \mu_m + \sum_{j=m}^{m+l-2}\alpha_j + \beta_{m+l-1} \right) \label{bound_Rml_2}.
\end{eqnarray}

Combining (\ref{bound_Rml_1}) and (\ref{bound_Rml_2}) gives the upper bound in (\ref{outer_Rml}).

Third, the first upper bound in (\ref{outer_Rsum}) is in fact the
non-symmetric version of \cite[Theorem 2]{zhou_kuser}, from which we have
\begin{eqnarray}
R_{sum} -  n \epsilon_n &\le& \sum _{k=1}^{K}\{ h(y_{ki}|s_{ki}) - h(z_{ki})\}  \nonumber  \\
&\le& n\sum_{j=1}^{K} \alpha_j \label{bound_Rsum_1}.
\end{eqnarray}
The other sum-rate upper bounds (i.e., $\rho_l$) can be derived by giving genies $x_l^n$ to $y^n_{l-1}$ and $s_j^n$ to $y_j^n$ for $j=1, 2, \cdots, K, j \neq l,l-1 $:
\begin{eqnarray}
\lefteqn{n(R_{sum} - \epsilon_n)} \nonumber \\
 &\le& I(x_1^n; y_1^n) + I(x_2^n; y_2^n) + \cdots + I(x_K^n; y_K^n) \nonumber \\
&=& I(x_{l-1}^n; y_{l-1}^n |x_{l}^n) + I(x_l^n; y_l^n) + \sum_{j=1, j\neq l, l-1}^{K} I(x_j^n; y_j^n s_j^n)  \nonumber \\
&=& h(h_{l-1, l-1}x_{l-1}^n + z_{l-1}^n) - h(z_{l-1}^n) + h(y_l^n) - h(s_{l+1}^n) \nonumber \\
&& + \sum_{j=1, j\neq l, l-1}^{K} \left[ h(s_j^n) -h(z_{j-1}^n) + h(y_j^n|s_j^n) - h(s_{j+1}^n) \right] \nonumber \\
&=& h(y_l^n) - h(z_{l-1}^n) + h(h_{l-1, l-1}x_{l-1}^n + z_{l-1}^n) \nonumber \\
&&  - h(h_{l-1, l-2}x_{l-1}^n + z_{l-2}^n)  \nonumber \\
&&  +\sum_{j=1, j\neq l, l-1}^{K} \left[ h(y_j^n|s_j^n) - h(z_{j-1}^n)  \right]   \nonumber \\
&\le& n \left( \beta_{l-1} + \gamma_l + \sum_{j=1, j\neq l, l-1}^{K} \alpha_j   \right) \nonumber \\
&=& n\rho_l
\end{eqnarray}
where $l=1, 2, \cdots, K$.

Fourth, for the bound in (\ref{outer_Rsi}), from Fano's inequality, we have
\begin{eqnarray}
\lefteqn{n(R_{sum} + R_i - \epsilon_n)}  \nonumber \\
&\le& \sum_{j=1}^{K}I(x_j^n; y_j^n)  + I(x_i^n; y_i^n) \nonumber \\
&\stackrel{(a)}{\le}& I(x_{i}^n; y_{i}^n)  +  I(x_{i}^n; y_{i}^n | x_{i+1}^{n}) + \sum_{j=1, j \neq i}^{K}I(x_j^n; y_j^n s_j^n)  \nonumber  \\
&=& h(y_i^n) - h(s_{i+1}^n) + h(h_{i,i}x_i^n + z_i^n) - h(z_{i}^n) \nonumber \\
&&  +\sum_{j=1, j \neq i}^{K} \left[ h(s_j^n) -h(z_{j-1}^n) + h(y_j^n|s_j^n) - h(s_{j+1}^n) \right]   \nonumber \\
&=& h(y_i^n) - h(z_i^n) +  h(h_{i,i}x_i^n + z_i^n) - h(h_{i,i-1}x_i^n + z_i^n) \nonumber \\
&& + \sum_{j=1, j \neq i}^{K} \left[ h(y_j^n|s_j^n) - h(z_{j-1}^n) \right] \nonumber \\
&\le& n \left( \beta_i + \gamma_i +\sum_{j=1, j \neq i }^{K} \alpha_j \right)
\end{eqnarray}
where in (a) we give genie $x_{i+1}^n$ to $y_{i}^n$ and $s_j^n$ to $y_j^n$ for $j=1, 2, \cdots, K, j \neq i$.

\subsection{Proof of $\mathcal{R}_{\mathrm{HK-TS}}^{(3)} \subseteq \mathcal{R}_{\mathrm{HK}}^{(3)}$ }
\label{R_NEW_R_HK_inclusion}

For a fixed $P_3 \subseteq \mathcal{P}_3$, define
\begin{equation}
P_3^* = \sum_{w_1}P_3, \quad P_3^{**} = \sum_{w_2}P_3, \quad P_3^{***} = \sum_{w_3}P_3.
\end{equation}
We will show that
\begin{eqnarray} \label{R_new_inclusiong}
\lefteqn{\mathcal{R}_{\mathrm{HK-TS}}^{(3)}(P_3)} \\
&& \subseteq \mathcal{R}_{\mathrm{HK}}^{(3)}(P_3) \cup  \mathcal{R}_{\mathrm{HK}}^{(3)}(P_3^{*}) \cup  \mathcal{R}_{\mathrm{HK}}^{(3)}(P_3^{**}) \cup  \mathcal{R}_{\mathrm{HK}}^{(3)}(P_3^{***}). \nonumber
\end{eqnarray}

Suppose that rate triple
$(R_1, R_2, R_3)$ is in $\mathcal{R}_{\textrm{HK-TS}}^{(3)}(P_3)$ but not
in $\mathcal{R}_{\textrm{HK}}^{(3)}(P_3)$.
Then at least one of the following inequalities hold:
\begin{eqnarray}
a_1 + e_3 \le R_1 \le d_1,  \label{R1_violated} \\
a_2 + e_1 \le R_2 \le d_2,  \label{R2_violated} \\
a_3 + e_2 \le R_3 \le d_3,  \label{R3_violated}
\end{eqnarray}
Without loss of generality, assume that (\ref{R1_violated}) holds.

Substituting $W_1 = \emptyset$ into $\mathcal{R}_{\mathrm{HK}}^{(3)}(P_3)$, we obtain $\mathcal{R}_{\mathrm{HK}}^{(3)}(P_3^{*})$ as follows:
\begin{eqnarray}
R_1 &\le& d_1, \\
R_2 &\le& \min\{d_2, a_2 + g_1 \}, \\
R_3 &\le& \min\{I(Y_3; X_3|Q), \nonumber \\
&&  e_2 + I(Y_3;X_3|W_3, Q)   \}, \\
R_1 + R_2 &\le& a_2 + g_1, \\
R_2 + R_3 &\le& \min \{g_2 + I(Y_3;X_3|W_3, Q), \nonumber \\
&& g_1 + e_2 + I(Y_3; X_3|W_3,Q)  \}, \\
R_3 + R_1 &\le& \min \{d_1 + I(Y_3; X_3|Q), \nonumber \\
&&  d_1 + e_2 + I(Y_3; X_3|W_3, Q)  \}, \\
R_1 + R_2 + R_3 &\le& g_1 + e_2 + I(Y_3; X_3|W_3, Q).
\end{eqnarray}
We will show that whenever $(\ref{R1_violated})$ is true, we have
$\mathcal{R}_{\mathrm{HK-TS}}^{(3)}(P_3) \subseteq \mathcal{R}_{\mathrm{HK}}^{(3)}(P_3^{*})$.
To this end, inspect $\mathcal{R}_{\mathrm{HK-TS}}^{(3)}(P_3)$ in
(\ref{R_new_R})-(\ref{R_new_Rsum+R3}). From (\ref{R_new_R}), we have
\begin{equation}
R_1 \le d_1, \label{R_new_P3_start}
\end{equation}
and from (\ref{R_new_R}) and (\ref{R1_violated}) and (\ref{R_new_R1+R2}), we have
\begin{eqnarray}
R_2 &\le& \min\{ d_2, a_2 + e_1 - a_1 \}  \nonumber \\
&\le&  \min\{ d_2, a_2 + g_1 \},
\end{eqnarray}
and from (\ref{R1_violated}) and (\ref{R_new_R3+R1}), we have
\begin{eqnarray}
R_3 &\le& \min\{g_3 - e_3, e_2  \}  \nonumber \\
&\le& \min\{ I(Y_3; X_3|Q), e_2 + I(Y_3;X_3|W_3, Q) \},
\end{eqnarray}
and from (\ref{R_new_R1+R2}), we have
\begin{equation}
R_1 + R_2 \le a_2 + g_1,
\end{equation}
and from (\ref{R1_violated}) and (\ref{R_new_R1+R2+R3}), we have
\begin{eqnarray}
R_2 + R_3 &\le& \min\{g_2, e_1 + e_2 - a_1 \}  \nonumber \\
&\le& \min \{g_2 + I(Y_3;X_3|W_3, Q), \nonumber \\
&& g_1 + e_2 + I(Y_3; X_3|W_3,Q)  \},
\end{eqnarray}
and from (\ref{R1_violated}) and (\ref{R_new_R3+R1}), we have
\begin{eqnarray}
R_3+R_1 &\le& \min\{d_1 + g_3 -a_3, e_2+d_1  \} \nonumber \\
&\le& \min \{d_1 + I(Y_3; X_3|Q), \nonumber \\
&& d_1 + e_2 + I(Y_3; X_3|W_3, Q)  \},
\end{eqnarray}
and from (\ref{R1_violated}) and (\ref{R_new_Rsum+R1}), we have
\begin{eqnarray}
R_1 + R_2 +R_3 &\le& g_1 + e_2 \nonumber \\
&\le& g_1 + e_2 + I(Y_3; X_3|W_3, Q). \label{R_new_P3_end}
\end{eqnarray}

It is  easy to see that $(R_1, R_2, R_3)$ satisfying the above constrains
(\ref{R_new_P3_start})-(\ref{R_new_P3_end}) is within the rate region $\mathcal{R}_{\mathrm{HK}}^{(3)}(P_3^{*})$.
In the same way, we can prove the cases for when (\ref{R2_violated}) holds and when (\ref{R3_violated}) holds.

Therefore, (\ref{R_new_inclusiong}) is true, and it immediately follows that
\begin{equation}
\mathcal{R}_{\mathrm{HK-TS}}^{(3)} \subseteq \mathcal{R}_{\mathrm{HK}}^{(3)}.
\end{equation}

\subsection{Useful Inequalities}
\label{useful_ineq}
Keep in mind that, with the ETW's power splitting strategy, i.e., $\mathsf{SNR}_{ip}=\min\{\mathsf{SNR}_i, \frac{\mathsf{SNR}_i}{\mathsf{INR}_i} \}$, we always have $\mathsf{SNR}_{ip} > \frac{\mathsf{SNR}_i}{1 + \mathsf{INR}_i}$. This appendix presents several useful inequalities as follows. For all $i=1, 2, \cdots, K$,
\begin{itemize}
\item $\lambda_i - d_i < 1$, because
\begin{eqnarray}
\lambda_i - d_i&=& \log(1+\mathsf{SNR}_i) - \log(2+\mathsf{SNR}_{i}) + 1  \nonumber \\
&=& 1 - \log \left( \frac{2 +\mathsf{SNR}_i }{1+\mathsf{SNR}_i} \right) \nonumber \\
&\le& 1
\end{eqnarray}

\item $\lambda_i - (a_i + e_{i-1}) < 2$, because
\begin{eqnarray}
\lefteqn{\lambda_i - (a_i + e_{i-1})} \nonumber \\
&=& \log(1+\mathsf{SNR}_i) - \log \left( 2+ \mathsf{SNR}_{ip} \right) +1 \nonumber \\
&& - \log \left( 1 + \mathsf{INR}_{i} + \mathsf{SNR}_{i-1, p}\right) + 1  \nonumber \\
& < &  2 + \log(1+\mathsf{SNR}_i) - \log \left( 1+ \frac{\mathsf{SNR}_i}{1 + \mathsf{INR}_i} \right) \nonumber \\
&& - \log \left( 1 + \mathsf{INR}_{i} \right) \nonumber \\
&=& 2 - \log \left(1 +  \frac{\mathsf{INR}_i}{1 + \mathsf{SNR}_i} \right) \nonumber \\
&\le& 2
\end{eqnarray}

\item $\beta_i - a_i < 1$, because
\begin{eqnarray}
\lefteqn{\beta_i - a_i} \nonumber \\
&=& \log \left( \frac{1+\mathsf{SNR}_i}{1+\mathsf{INR}_i} \right) -   \log \left( 2+ \mathsf{SNR}_{ip} \right) + 1 \nonumber \\
&<& \log \left( \frac{1+\mathsf{SNR}_i}{1+\mathsf{INR}_i} \right)  - \log \left(1 + \frac{\mathsf{SNR}_i}{1 + \mathsf{INR}_i} \right) +1 \nonumber \\
&=& 1 - \log \left(1 +  \frac{\mathsf{INR}_i}{1 + \mathsf{SNR}_i} \right) \nonumber \\
&\le& 1
\end{eqnarray}

\item $\alpha_i - e_i < 1$, because
\begin{eqnarray}
\alpha_i - e_i &=& \log \left(1+ \mathsf{INR}_{i+1} + \frac{\mathsf{SNR}_i}{1+\mathsf{INR}_i} \right) \nonumber \\
&& - \log \left( 1 + \mathsf{INR}_{i+1} + \mathsf{SNR}_{ip} \right) + 1   \nonumber\\
&\le& 1
\end{eqnarray}

\item $\gamma_i - g_i = 1$, because
\begin{eqnarray}
\gamma_i - g_i &=& \log \left( 1 + \mathsf{INR}_{i+1} + \mathsf{SNR}_i \right) \nonumber \\
&& - \log \left( 1 + \mathsf{INR}_{i+1} + \mathsf{SNR}_i \right)  + 1 \nonumber\\
&=&1
\end{eqnarray}

\item $\mu_i - e_{i-1} < 1$, because
\begin{eqnarray}
\mu_i - e_{i-1} &=& \log (1 + \mathsf{INR}_i) \nonumber \\
&& - \log \left( 1 + \mathsf{INR}_{i} + \mathsf{SNR}_{i-1,p} \right) + 1  \nonumber \\
&\le& 1
\end{eqnarray}
\end{itemize}

\bibliographystyle{IEEEtran}


\begin{IEEEbiographynophoto}{Lei Zhou}
(S'05) received the B.E. degree in electronics engineering
from Tsinghua University, Beijing, China,
in 2003 and M.A.Sc. degree in electrical and computer engineering from the University of Toronto, ON, Canada, in 2008. During 2008-2009, he was with Nortel Networks, Ottawa, ON, Canada. He is currently pursuing the Ph.D. degree with the Department of Electrical and Computer Engineering, University of Toronto, Canada. His research interests include multiterminal information theory, wireless communications, and signal processing.

He is a recipient of the Shahid U.H. Qureshi Memorial Scholarship in 2011, the Alexander Graham Bell Canada Graduate Scholarship for 2011-2013, and the Chinese government award for outstanding self-financed students abroad in 2012.

\end{IEEEbiographynophoto}

\begin{IEEEbiographynophoto}{Wei Yu}
(S'97-M'02-SM'08) received the B.A.Sc. degree in Computer Engineering and
Mathematics from the University of Waterloo, Waterloo, Ontario, Canada in 1997
and M.S. and Ph.D. degrees in Electrical Engineering from Stanford University,
Stanford, CA, in 1998 and 2002, respectively. Since 2002, he has been with the
Electrical and Computer Engineering Department at the University of Toronto,
Toronto, Ontario, Canada, where he is now Professor and holds a
Canada Research Chair in Information Theory and Digital Communications. His
main research interests include multiuser information theory, optimization,
wireless communications and broadband access networks.

Prof. Wei Yu currently serves as an Associate Editor for
{\sc IEEE Transactions on
Information Theory}. He was an Editor for {\sc IEEE Transactions on Communications} (2009-2011), an Editor for {\sc IEEE Transactions on Wireless
Communications} (2004-2007), and a Guest Editor for a number of
special issues for the {\sc IEEE Journal on Selected Areas in
Communications} and the {\sc EURASIP Journal on Applied Signal Processing}.
He is member of the Signal Processing for Communications and Networking
Technical Committee of the IEEE Signal Processing Society.
He received the IEEE Signal Processing Society Best Paper Award in 2008,
the McCharles Prize for Early Career Research Distinction in 2008,
the Early Career Teaching Award from the Faculty of Applied Science
and Engineering, University of Toronto in 2007, and the Early Researcher
Award from Ontario in 2006.
\end{IEEEbiographynophoto}

\end{document}